# SOURCE CODE RETRIEVAL USING SEQUENCE BASED SIMILARITY

Yoshihisa Udagawa

Faculty of Engineering, Tokyo Polytechnic University, Atsugi City, Kanagawa, Japan
udagawa@cs.t-kougei.ac.jp

### ABSTRACT

*Duplicate code adversely affects the quality of software systems and hence should be detected. We discuss an approach that improves source code retrieval using structural information of source code. A lexical parser is developed to extract control statements and method identifiers from Java programs. We propose a similarity measure that is defined by the ratio of the number of sequential fully matching statements to the number of sequential partially matching statements. The defined similarity measure is an extension of the set-based Sorensen-Dice similarity index. This research primarily contributes to the development of a similarity retrieval algorithm that derives meaningful search conditions from a given sequence, and then performs retrieval using all derived conditions. Experiments show that our retrieval model shows an improvement of up to 90.9% over other retrieval models relative to the number of retrieved methods.*

## 1. INTRODUCTION

Several studies have shown that approximately 5%–20% of a program can contain duplicate code [2, 13]. Many such duplications are often the result of copy-paste operations, which are simple and can significantly reduce programming time and effort when the same functionality is required.

In many cases, duplicate code causes an adverse effect on the quality of software systems, particularly the maintainability and comprehensibility of source code. For example, duplicate code increases the probability of update anomalies. If a bug is found in a code fragment, all the similar code fragments should be investigated to fix the bug in question [11, 15]. This coding practice also produces code that is difficult to maintain and understand, primarily because it is difficult for maintenance engineers to determine which fragment is the original one and whether the copied fragment is intentional. Tool support that efficiently and effectively retrieves similar code is required to support software engineers' activities.

Different approaches for identifying similar code fragments have been proposed in code clone detection. Based on the level of analysis applied to the source code, clone detection techniques can be roughly classified into four main groups, i.e., text-based, token-based, structure-based, and metrics-based.

### (1) Text-based approaches

In this approach, the target source program is considered as a sequence of strings. Baker [2] described an approach that identifies all pairs of matching "parameterized" code fragments. Johnson [7] proposed an approach to extract repetitions of text and a matching mechanism using fingerprints on a substring of the source code. Although these methods achieve high performance, they are sensitive to lexical aspects, such as the presence or absence of new lines and the ordering of matching lines.

### (2) Token-based approaches

In the token-based detection approach, the entire source system is transformed into a sequence of tokens, which is then analyzed to identify duplicate subsequences. A sub-string matching algorithm is generally used to find common subsequences. CCFinder [22] adopts the token-based technique to efficiently detect "copy and paste" code clones. In CCFinder, the similarity metric between two sets of source code files is defined based on the concept of "correspondence." CP-Miner [11] uses a frequent subsequence mining technique to identify a similar sequence of tokenized statements. Token-based approaches are typically more robust against code changes compared to text-based approaches.

### (3) Structure-based approaches

In this approach, a program is parsed into an abstract syntax tree (AST) or program dependency graph (PDG). Because ASTs and PDGs contain structural information about the source code, sophisticated methods can be applied to ASTs and PDGs for the clone detection. CloneDR [3] is one of the pioneering AST-based clone techniques. Wahler et al. [21] applied frequent itemset data mining techniques to ASTs represented in XML to detect clones with minor changes. DECKARD [6] also employs a tree-based approach in which certain characteristic vectors are computed to approximate the structural information within ASTs in Euclidean space.

Typically, a PDG is defined to contain the control flow and data flow information of a program. An isomorphic subgraph matching algorithm is applied to identify similar subgraphs. Komondoor et al. [8] have also proposed a tool for C programs that finds clones. They use PDGs and a program slicing technique to find clones. Krinke [10] uses an iterative approach (k-length patch matching) to determine maximal similar subgraphs. Structure-based approaches are generally robust to code changes, such as reordered, inserted, and deleted code. However, they are not scalable to large programs.

### (4) Metrics-based approaches

Metrics-based approaches calculate metrics from code fragments and compare these metric vectors rather than directly comparing source code. Kontogiannis et al. [9] developed an abstract pattern matching tool to measure similarity between two programs using Markov models. Some common metrics in this approach include a set of software metrics called "fingerprinting" [7], a set of method-level metrics including McCabe's cyclomatic complexity [14], and a characteristic vector to approximate the structural information in ASTs [6].

Our approach is classified as a structure-based comparison. It features a sequence of statements as a retrieval condition. We have developed a lexical parser to extract source code structure, including control statements and method identifiers. The extracted structural information is input to a vector space model [1,12,17], an extended Sorensen-Dice model [4,16,19], and the proposed source code retrieval model, named the "derived sequence retrieval model" (DSRM). The DSRM takes a sequence of statements as a retrieval condition and derives meaningful search conditions from the given sequence. Because a program is composed of a sequence of statements, our retrieval model improves the performance of source code retrieval.

The remainder of this paper is organized as follows. In Section 2, we present an outline of the process and the target source code of our research. In Section 3, we define source code similarity metrics. Retrieval results are discussed in Section 4. In Section 5, we analyze performance in elapsed time, and Section 6 presents conclusions and suggestions for future work.

## 2. RESEARCH PROCESS

### 2.1. Outline

Figure 1 shows an outline of our research process. Generally, similarity retrieval of source code is performed for a specific purpose. From this perspective, the original source code may include some uninteresting fragments. We have developed a lexical parser and applied it to a set of original Java source codes to extract interesting code, which includes class method signatures, control statements, and method calls. Our parser traces a variable type declaration and class instantiation to generate an identifier-type list. This list is then used to translate a variable identifier to its data type. A method call preceded by an identifier is converted into the method calls preceded by the data type of the identifier.

Code matching is performed using three retrieval models. The first model is the proposed DSRM, which takes a sequence of statements as a retrieval condition. The second model is based on the collection of statements, and is referred to as the derived collection retrieval model (DCRM). The DCRM is an extension of the Sorensen-Dice model of index [4,16,19]. The final retrieval model is the vector space model (VSM) [1,12,17], which has been developed to retrieve a natural language document. Source code can be perceived as a highly structured document; therefore, comparison with DSRM, DCRM, and VSM provides a baseline performance evaluation of how structure of a document will affect retrieval results.

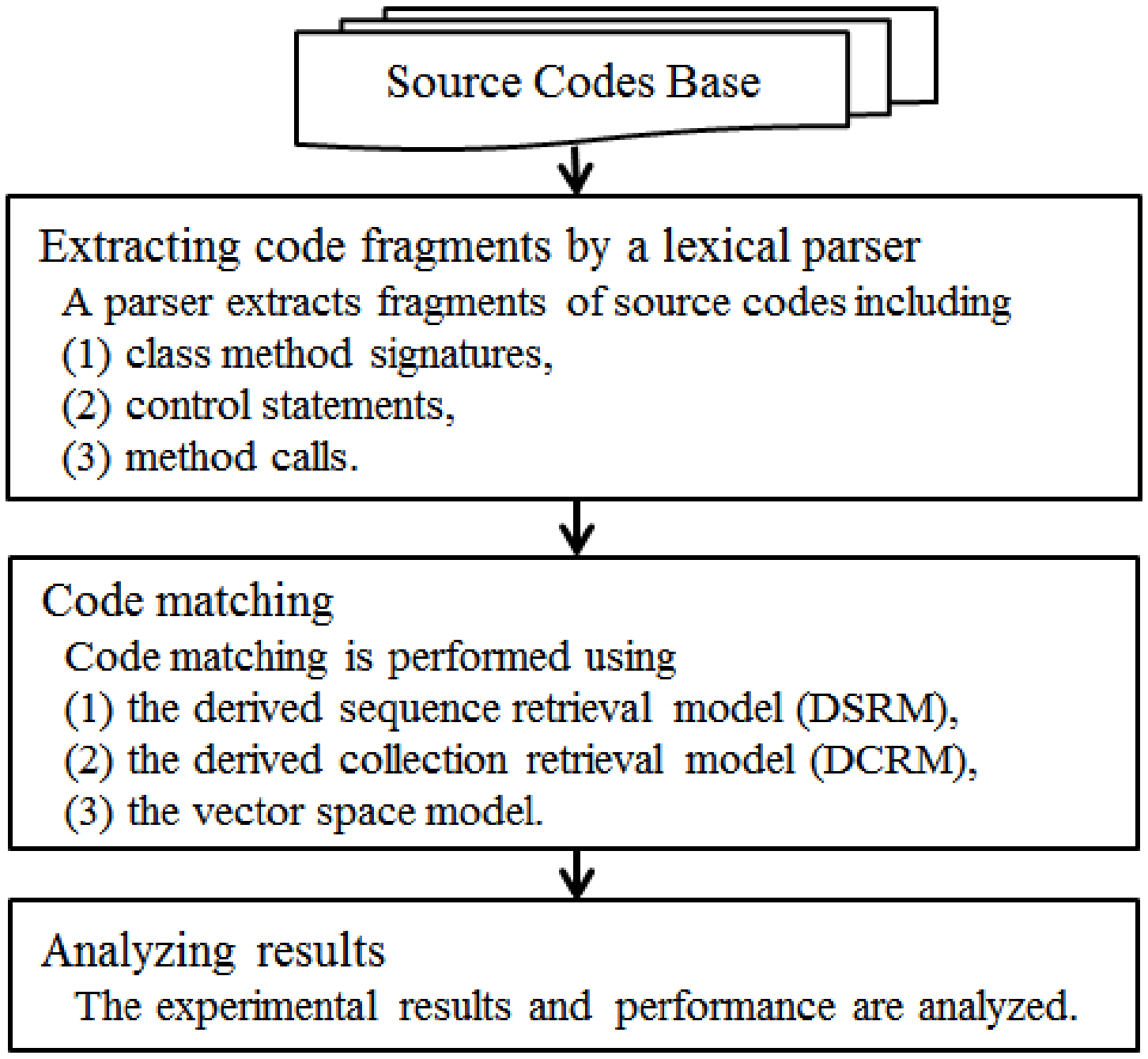

Figure 1. Outline of our research process

## 2.2. Extracting Source Code Segments

At the beginning of our approach, a set of Java source codes is partitioned into methods. Then, the code matching statements are extracted for each method. The extracted fragments comprise class method signatures, control statements, and method calls.

(1) Class method signatures

Each method in Java is declared in a class [5]. Our parser extracts class method signatures in the following syntax.

<class identifier>:<method signature>

An anonymous class, which is a local class without a class declaration, is often used when a local class is used only once. An anonymous class is defined and instantiated in a single expression using the new operator to make code concise. Our parser extracts a method declared in an anonymous class in the following syntax.

<class identifier>:<anonymous class identifier>:<method signature>

Arrays and generics are widely used in Java to facilitate the manipulation of data collections. Our parser also extracts arrays and generic data types according to Java syntax. For example, Object[], String[][], List<String>, and List<Integer> are extracted and treated as different data types.

(2) Control statements

Our parser also extracts control statements with various levels of nesting. The block is represented by the "{" and "}" symbols. Hence, the number of "{" symbols indicates the number of nesting levels. The following Java control statements [5] are extracted by our parser.

- If-then (with or without else or else if statements)
- Switch
- While
- Do
- For and enhanced for
- Break
- Continue
- Return
- Throw
- Synchronized
- Try (with or without a catch or finally clause)

(3) Method calls

From the assumption that a method call characterizes a program, our parser extracts a method identifier called in a Java program. Generally, the instance method is preceded by a variable whose type refers to a class object to which the method belongs. Our parser traces a type declaration of a variable and translates a variable identifier to its data type or class identifier, i.e.,

<variable>.<method identifier>

is translated into

<data type>.<method identifier>
or
<class identifier>.<method identifier>.

## 2.3. Extracting Statements of Struts 2

We selected Struts 2.3.1.1 Core as our target because Struts 2 [20] is a popular Java framework for web applications. We estimated the volume of source code using file metrics. Typical file metrics are as follows:

| | | |
|---|---|---|
| Java Files | ---- | 368 |
| Classes | ---- | 414 |
| Methods | ---- | 2,667 |
| Lines of Code | ---- | 21,543 |
| Comment Lines | ---- | 17,954 |
| Total Lines | ---- | 46,100 |

Struts 2.3.1.1 Core consists of 46,100 lines of source code, including blank and comment lines. Struts 2.3.1.1 Core is classified as mid-scale software in the industry. Struts 2.3.1.1 Core is comprised of 368 Java files, which differs from the number of declared classes (414) because some Java files include definitions of inner classes and anonymous classes. Figure 2 shows an example of the extracted structure of the *evaluateClientSideJsEnablement()* method in the *Form.java* file of the *org.apache.struts2.components* package. The three numbers preceded by the # symbol are the number of comment, blank, and code lines, respectively. The extracted structures include the depth of nesting of the control statements; thus, they supply sufficient information for retrieving methods using a source code substructure.

```
Form::void evaluateClientSideJsEnablement(String, String, String)
#       8       0       22
{
(
        if{
        addParameter
        Configuration.getRuntimeConfiguration
        RuntimeConfiguration.getActionConfig
                if{
                ActionConfig.getInterceptors
                        for{
                                if{
                                interceptorMapping.getInterceptor
                                ValidationInterceptor.getExcludeMethodsSet
                                ValidationInterceptor.getIncludeMethodsSet
                                        if{
                                        addParameter
                                        }
                                }
                        }
                }
        }
}
```

Figure 2. Example of extracted structure

# 3. SIMILARITY METRICS

## 3.1. Vector Space Model for Documents

The VSM is widely used in retrieving and ranking documents written in natural languages. Documents and queries are represented as vectors. Each dimension of the vectors corresponds to a term that consists of documents and queries. The documents are ranked against queries by computing similarity, which is computed as the cosine of the angle between the two vectors.
Given a set of documents $D$, a document $\boldsymbol{d}_j$ in $D$ is represented as a vector of term weights:

$$\boldsymbol{d_j} = (w_{1,j}, w_{2,j}, \dots, w_{N,j})$$

where $N$ is the total number of terms in document $\boldsymbol{d}_j$, and $w_{i,j}$ is the weight of the $i$-th term.

There are many variations of the term weighting scheme. Salton et al. [17] proposed the well-known “term frequency-inverse document frequency” (tf-idf) weighting scheme. According to this weighting scheme, the weight of the $j$-th element of the document $\boldsymbol{d}_j$, i.e., $w_{i,j}$, is computed by the product of the term frequency $tf_{i,j}$ and the inverse document frequency $idf_i$.

$$w_{i,j} = tf_{i,j} \cdot idf_i$$

The term frequency $tf_{i,j}$ is defined as the number of occurrences of the term $i$ in the document $\boldsymbol{d}_j$. The inverse document frequency is a measure of the general importance of the term $i$ and is defined as follows:

$$idf_i = log_2 \left[\frac{M}{df_i}\right]$$

where M denotes the total number of documents in a collection of documents. A high weight in $w_{i,j}$ is reached by a high term frequency in the given document and a low document frequency $df_i$ of the term in the whole collection of documents. Hence, the weights tend to filter out common terms.

A user query can be similarly converted into a vector $\boldsymbol{q}$:

$$\boldsymbol{q} = \left(w_{1,q}, w_{2,q}, \ldots, w_{N,q}\right)$$

The similarity between document $\boldsymbol{d}_j$ and query $\boldsymbol{q}$ can be computed as the cosine of the angle between the two vectors $\boldsymbol{d}_j$ and $\boldsymbol{q}$ in the $N$-dimensional term space:

$$Sim_{cos}(\boldsymbol{d_j}, \boldsymbol{q}) = \frac{\sum_{i=1}^{N} w_{i,j} * w_{i,q}}{\sqrt{\sum_{i=1}^{N} w_{i,j}^2} * \sqrt{\sum_{i=1}^{N} w_{i,q}^2}}$$

This similarity is often referred to as the cosine similarity.

## 3.2. Extending Sorensen-Dice Index

Over the last decade, many techniques that detect software cloning and refactoring opportunities have been proposed. Similarity coefficients play an important role in the literature. However, most similarity definitions are validated by empirical studies. The choice of measure depends on the characteristics of the domain to which they are applied. Among many different similarity indexes, the similarity defined in CloneDR is worth notice. Baxter et al. [3] define the similarity between two trees T1 and T2 as follows:

Similarity(T1, T2) = 2H / (2H + L + R)

where H is the number of shared nodes in trees T1 and T2, L is the number of unique nodes in T1, and R is the number of unique nodes in T2. Within the context of a tree structure, this definition can be seen as an extension of the Sorensen-Dice index.

The Sorensen-Dice index is originally defined by two sets and is formulated as follows:

$$\mathrm{Sim}_{\text{Sorensen-Dice}}(X_1, X_2) = \frac{2|X_1 \cap X_2|}{2|X_1 \cap X_2| + |X_1 \cap \neg X_2| + |\neg X_1 \cap X_2|}$$

Here, $|X_1 \cap X_2|$ indicates the number of elements in the intersection of sets $X_1$ and $X_2$.

Another well-known index is the Jaccard index of binary features, which is defined by the following formula:

$$\mathrm{Sim}_{\mathrm{Jaccard}}(X_1, X_2) = \frac{|X_1 \cap X_2|}{|X_1 \cap X_2| + |X_1 \cap \neg X_2| + |\neg X_1 \cap X_2|}$$

In software, the Sorensen-Dice index and the Jaccard index are known experimentally to produce better results than other indexes, such as a simple matching index, which counts the number of features absent in both sets [16,19]. The absence of a feature in two entities does not indicate similarity in software source code. For example, if two classes do not include the same method, it does not mean that the two classes are similar. The Jaccard and Sorensen-Dice indexes perform identically except for the value of the similarity because assigning more weight to the features present in both entities does not have a significant impact on the results. Our study takes the Sorensen-Dice index as a basis for defining the similarity measure between source codes. The extension of the Sorensen-Dice index on N sets is straightforward.

$$\mathrm{Sim}_{\mathrm{Sorensen\text{-}Dice}}(X_1, X_2, \dots, X_n) = \frac{n\,|X_1 \cap X_2 \dots \cap X_n|}{\sum_{r=0}^{r=n-1} (n-r)\,|\,SetComb(X_1 \cap X_2 \dots \cap X_n,\ r)\,|}$$

The function $SetComb(X_1 \cap X_2 \cap \dots \cap X_n, r)$ defines intersections of sets $\{X_1, X_2, \dots, X_n\}$ whose r elements are replaced by the elements with the negation symbol. The summation of $r = 0$ to $n-1$ of $SetComb(X_1 \cap X_2 \cap \dots \cap X_n, r)$ generates the power set of sets $X_1, X_2, \dots, X_n$, excluding the empty set. $(n-r)$ indicates the number of sets without the negation symbol. $|X_1 \cap X_2, \dots, \cap X_n|$ indicates the number of tuples $<x_1, x_2, \dots, x_n>$ where $x_1 \in X_1, x_2 \in X_2, \dots, x_n \in X_n$.

For example, in case $n = 3$, the numerator of the extended Sorensen-Dice index on sets $X_1$, $X_2$, and $X_3$ equals $3|X_1 \cap X_2 \cap X_3|$, and the denominator equals $3|X_1 \cap X_2 \cap X_3| + 2|X_1 \cap X_2 \cap \neg X_3| + 2|X_1 \cap \neg X_2 \cap X_3| + 2|\neg X_1 \cap X_2 \cap X_3| + |X_1 \cap \neg X_2 \cap \neg X_3| + |\neg X_1 \cap X_2 \cap \neg X_3| + |\neg X_1 \cap \neg X_2 \cap X_3|$.

## 3.3. Similarity Metric for Source Codes

In the vector space retrieval model, a document is represented as a vector of terms that comprise the document. The similarity of a document and a query is calculated as the cosine of the angle between a document vector and a query vector. This means that the order in which the terms appear in a document is lost in the vector space model. On the other hand, a computer program is a sequence of instructions written to perform a specified task [18]. The source code is essentially a sequence of characters forming a more complex text structure, such as statements, blocks, classes, and methods. This means that it is preferable or even crucial to consider the order of terms for a similarity index. In our study, the similarity measure is tailored to measure the similarity of sequentially structured text.

We first define the notion of a sequence. Let $S_1$ and $S_2$ be statements extracted by the structure extraction tool. $[S_1 \rightarrow S_2]$ denotes a sequence of $S_1$ followed by $S_2$. In general, for a positive integer n, let $S_i$ (i ranges between 1 and n) be a statement. $[S_1 \rightarrow S_2 \rightarrow \dots \rightarrow S_n]$ denotes a sequence of n statements.

The similarity of the DSRM can be considered the same as the extended Sorensen-Dice index except for symbols, i.e., using → symbol in place of ∩ symbol. The DSRM's similarity between two sequences is defined as follows:

$$\mathrm{Sim}_{DSRM}\,(\,[S_1\rightarrow S_2\rightarrow\ldots\rightarrow S_m],\,[T_1\rightarrow T_2\rightarrow\ldots\rightarrow T_n]\,) = \frac{n\,|\,[S_1\rightarrow S_2\rightarrow\cdots\rightarrow S_m],\ [T_1\rightarrow T_2\rightarrow\cdots\rightarrow T_n]\,|}{\sum_{r=0}^{r=n-1}(n-r)\,\|\,[S_1\rightarrow S_2\rightarrow\cdots\rightarrow S_m],\ SqcComb([T_1\rightarrow T_2\rightarrow\cdots\rightarrow T_n],\ r)\,\|}$$

Here, without loss of generality, we can assume that $m \geq n$. In case $m < n$, we replace the sequence $[S_1\rightarrow S_2 \rightarrow\ldots\rightarrow S_m]$ with $[T_1\rightarrow T_2\rightarrow\ldots\rightarrow T_n]$.

The numerator of the definition, i.e., $|\,[S_1\rightarrow S_2 \rightarrow\ldots\rightarrow S_m],\ [T_1\rightarrow T_2\rightarrow\ldots\rightarrow T_n]\,|$ indicates the number of statements in the sequence where $S_{j+1}=T_1$, $S_{j+2}=S_2$, ... , $S_{j+n}=T_n$ for some j ($0 \leq j \leq m - n$). The denominator of the definition indicates the iteration of the sequence match that counts the sequence of statements from $r = 0$ to $r = n-1$. Note that the first sequence $[S_1\rightarrow S_2 \rightarrow\ldots\rightarrow S_m]$ is renewed when the sequence match succeeds, i.e., replacing the matched statements with a not applicable symbol "n/a." *SqcComb* $([T_1\rightarrow T_2\rightarrow\ldots\rightarrow T_n],\ r)$ generates a set of sequence combinations by replacing the r ($0 \leq r < n$) statements with the negation of the statements.

For example, for $m = 4$ and $n = 2$, $\mathrm{Sim}_{DSRM}$ ( $[A_1\rightarrow A_1\rightarrow A_2\rightarrow A_2]$, $[A_1\rightarrow A_2]$ ) equals 0.5. The numerator of $\mathrm{Sim}_{DSRM}$ ( $[A_1\rightarrow A_1\rightarrow A_2\rightarrow A_2]$, $[A_1\rightarrow A_2]$ ) is 2 because the sequence $[A_1\rightarrow A_2]$ is included in the first sequence, $2*|\,[A_1\rightarrow A_1\rightarrow A_2\rightarrow A_2],\ [A_1\rightarrow A_2]\,| = 2*1 = 2$. The denominator of $\mathrm{Sim}_{DSRM}$ ($[A_1\rightarrow A_1\rightarrow A_2\rightarrow A_2]$, $[A_1\rightarrow A_2]$) is computed as follows. First, for set $r = 0$, *SqcComb*($[A_1\rightarrow A_2]$, 0) generates $[A_1\rightarrow A_2]$. Then, $2*\|\,[A_1\rightarrow A_1\rightarrow A_2\rightarrow A_2],\ [A_1\rightarrow A_2]\,\|$ is estimated as 2 because $[A_1\rightarrow A_2]$ is a subsequence of $[A_1\rightarrow A_1\rightarrow A_2\rightarrow A_2]$. Then, the first sequence is renewed by $[A_1\rightarrow n/a\rightarrow n/a\rightarrow A_2]$.

Next, for set $r = 1$, *SqcComb* ($[A_1\rightarrow A_2]$, 1) generates $[A_1\rightarrow\neg A_2]$ and $[\neg A_1\rightarrow A_2]$. $\|\,[A_1\rightarrow n/a\rightarrow n/a\rightarrow A_2],\ [A_1\rightarrow\neg A_2]\,\|$ is estimated as 1 because $A_1$ is included in $[A_1\rightarrow n/a\rightarrow n/a\rightarrow A_2]$, and then the first sequence is renewed by $[n/a\rightarrow n/a\rightarrow n/a\rightarrow A_2]$. Finally, $\|\,[n/a\rightarrow n/a\rightarrow n/a\rightarrow A_2],\ [\neg A_1\rightarrow A_2]\,\|$ is estimated as 1 and the first sequence is renewed by $[n/a\rightarrow n/a\rightarrow n/a\rightarrow n/a]$. The denominator of $\mathrm{Sim}_{DSRM}$ ($[A_1\rightarrow A_1\rightarrow A_2\rightarrow A_2]$, $[A_1\rightarrow A_2]$) is 4. Thus, $\mathrm{Sim}_{DSRM}$ ($[A_1\rightarrow A_1\rightarrow A_2\rightarrow A_2]$, $[A_1\rightarrow A_2]$ ) equals $2/4 = 0.5$.

```
// Datatype "method_structure" is a set of "sequences" of statements.
// A "sequence" is represented by an array of statements.
// Input: set_of_method_structure M;
// Input: sequence [S1→S2→...→Sn];
// Output: Sim[M.length];
// --- Definition of the SimDSRM function
double[] SimDSRM( set_of_method_structure M, sequence [S1→S2→...→Sn]){
  double Sim[M.length];     int Nume;     int Deno;
  for ( int j=0; j < M.length; j++ ) {
      Nume= Count( getMethodStructure(j), [ S1→S2→...→Sn], 0 );
      Deno= 0;
      for ( int r=1; r < [S1→S2→...→Sn].length; r++ ){
        Deno= Deno + Count( getMethodStructure(j), [ S1→S2→...→Sn], r );
      }
     Sim[j]= (double) Nume / (double)( Nume + Deno );
  }
  Return Sim;
}
// --- Definition of the Count function.
int Count( method_structure MS, sequence TN, int R){
  statement[] S;        // Type of S is a "sequence" of statements
  statement[][] DS;     // Type of DS is a set of a "sequence" of statements
  statement[] SV;       // Type of SV is a "sequence" of statements
  int CT=0;
  // Generate derived sequence replacing R statements with negations
  DS= SqcComb( TN, R );
  for ( each S[] ∈ MS){
    for ( int j=1; j<=MS.length–TN.length; j++ ){
      for ( each SV[] ∈ DS ){
        for ( int k=1; k<= TN.length–R; k++ ){
          if ( S[j] = SV[k] for all k-th elements that are not negative )
                  { CT= CT + (TN.length–R) };
        }
      }
    }
  }
 Return CT;
}
```

Figure 3. Algorithm to compute similarity for a sequence $[S_1 \rightarrow S_2 \rightarrow ... \rightarrow S_n]$

A simplified version of the algorithm to compute the DSRM's similarity is shown in Figure 3. It takes a set of method structures and a sequence as retrieval conditions as arguments and returns an array of similarity values for the set of method structures.

It is assumed that the *getMethodStructure(j)* function returns a structure of the j-th method extracted by the structure extraction tool. The function abstracts the implementation of the internal structure of the method. This is represented as a sequence of statements.

The Count function takes three arguments, i.e., a method_structure MS, a sequence of statements TN, and an integer R. Note that an element of the method_structure is compatible with a sequence of statements.

The SqcComb( TN, R ) function generates combinations of statement sequences that replace the R statements with the negation of the statements in the sequence TN. Then, matching between the method_structure MS and the combinations of statement sequences is processed. The Count function returns the number of positive statements that match the combinations of statement sequences.

The SimDSRM function calculates the similarity according to the DSRM's defined similarity. Note that the similarity is 1.0 when a method includes the sequence [$S_1 \rightarrow S_2 \rightarrow ... \rightarrow S_n$] and does not include any of the derived sequences from [$S_1 \rightarrow S_2 \rightarrow ... \rightarrow S_n$].

# 4. CODE RETRIEVAL EXPERIMENTS

## 4.1. Approach

Cosine similarity is extensively used in research on retrieving documents written in natural languages and recovering links between software artifacts [1,12]. Set-based indexes, such as the Jaccard index and the Sorensen-Dice index, are used in a variety of research, including software clustering [16] and generating refactoring guidelines [19]. Here, we present experimental results obtained using cosine similarity, the Sorensen-Dice index, and the DSRM's similarity.

## 4.2. Vector Space Model Results

It is natural to assign structural metrics to the elements of a document vector. For example, the *evaluateClientSideJs-Enablement()* method shown in Figure 2 is represented by the vector (4, 1, 2, 1, 1, 1, 1, 1, 1), where we assume that the first element of the vector corresponds to *if*-statements, the second corresponds to *for*-statements, the third corresponds to the *addParameter* method identifier, and the fourth corresponds to the *configuration.getRuntimeConfiguration* method identifier, and so on. Thus, the extracted fragments of Struts 2.3.1.1 Core are vectorized to produce a 1,420 × 2,667 matrix.

In Struts 2 Core, the *addParameter* method is often called after an *if*-statement because the *addParameter* method adds a key and a value given by the arguments to the parameter list maintained by the Struts 2 process after checking the existence of a key. Thus, the same number of *if*-statements and *addParameter* method identifiers are a reasonable retrieval condition in the vector space retrieval model.

Table 1 shows the top 27 methods retrieved by a query vector that consists of one *if*-statement, one *addParameter* method identifier, and one curly brace "}." The third column of Table 1 shows the similarity values calculated by the cosine similarity. It can be seen that 2,667 methods were retrieved because all methods include at least one curly brace "}." There were only 38 methods whose similarity values were greater than 0.3. The result looks fairly good at a glance; however, the results include some controversial methods in the sense that we are retrieving an *addParameter* method identifier that is called just after an *if*-statement. Figure 4 shows *ActionError::void evaluateExtraParams()*, which has the same structure as *Action Message::void evaluateExtraParams()* except for string arguments "actionErrors" and "actionMessages." The cosine similarity of *ActionError::void evaluateExtraParams()* is 0.846, and the extended Sorensen-Dice index is 0.750 because the method includes two *if*-statements and two *addParameter* methods. However, the method does not include any sequences of *if*-statements or an *addParameter* method. Thus, the DSRM's similarity is estimated to be 0.

Let a "boundary method" be a retrieved method whose DSRM's similarity is greater than 0 and whose cosine similarity is minimum. The *evaluateClientSideJsEnablement()*, which is shown at No.19 in Table 1, is the boundary method with the minimum cosine similarity 0.472. Table 1 consists of a set of retrieved methods whose cosine similarities are greater than or equal to the cosine similarity of the boundary method (0.472). The methods whose "No" column is shaded in Table 1 are those methods whose DSRM's similarities equals 0. The methods are a kind of controversial candidates. Details are discussed in the following sections.

**Table 1.** Top 27 retrieved methods

| No | Method Name | Vector Space Model + tfidf | Extended Sorensen -Dice Model | | | Derived Sequence Retrieval Model | | | Code Lines |
|---|---|---|---|---|---|---|---|---|---|
| | | Similarity | Similarity | Exact Match | Partial Match | Similarity | Exact Match | Partial Match | |
| 1 | TextArea::void evaluateExtraParams() | 0.967 | 0.923 | 12 | 1 | 0.923 | 12 | 1 | 15 |
| 2 | Form::void evaluateExtraParams() | 0.994 | 0.909 | 30 | 3 | 0.909 | 30 | 3 | 36 |
| 3 | TextField::void evaluateExtraParams() | 0.943 | 0.900 | 9 | 1 | 0.900 | 9 | 1 | 12 |
| 4 | DoubleListUIBean::void evaluateExtraParams() | 0.987 | 0.882 | 105 | 14 | 0.782 | 93 | 26 | 128 |
| 5 | UpDownSelect::void evaluateParams() | 0.907 | 0.774 | 24 | 7 | 0.774 | 24 | 7 | 41 |
| 6 | Anchor::void evaluateExtraParams() | 0.509 | 0.750 | 6 | 2 | 0.750 | 6 | 2 | 14 |
| 7 | Form::void populateComponentHtmlId(Form form) | 0.510 | 0.750 | 3 | 1 | 0.750 | 3 | 1 | 6 |
| 8 | Password::void evaluateExtraParams() | 0.686 | 0.750 | 3 | 1 | 0.750 | 3 | 1 | 6 |
| 9 | Reset::void evaluateExtraParams() | 0.686 | 0.750 | 3 | 1 | 0.750 | 3 | 1 | 5 |
| 10 | Submit::void evaluateExtraParams() | 0.686 | 0.750 | 3 | 1 | 0.750 | 3 | 1 | 5 |
| 11 | Select::void evaluateExtraParams() | 0.975 | 0.857 | 12 | 2 | 0.643 | 9 | 5 | 16 |
| 12 | ComboBox::void evaluateExtraParams() | 0.977 | 0.818 | 27 | 6 | 0.636 | 21 | 12 | 39 |
| 13 | OptionTransferSelect::void evaluateExtraParams() | 0.838 | 0.675 | 27 | 13 | 0.600 | 24 | 16 | 106 |
| 14 | FieldError::void evaluateExtraParams() | 0.858 | 0.600 | 3 | 2 | 0.600 | 3 | 2 | 6 |
| 15 | InputTransferSelect::void evaluateExtraParams() | 0.913 | 0.774 | 24 | 7 | 0.581 | 18 | 13 | 50 |
| 16 | Checkbox::void evaluateExtraParams() | 0.932 | 0.500 | 3 | 3 | 0.500 | 3 | 3 | 7 |
| 17 | Label::void evaluateExtraParams() | 0.747 | 0.692 | 9 | 4 | 0.462 | 6 | 7 | 18 |
| 18 | File::void evaluateParams() | 0.507 | 0.353 | 6 | 11 | 0.353 | 6 | 11 | 24 |
| 19 | Form::void evaluateClientSideJsEnablement() | 0.472 | 0.500 | 6 | 6 | 0.250 | 3 | 9 | 22 |
| 20 | ListUIBean::void evaluateExtraParams() | 0.811 | 0.727 | 24 | 9 | 0.182 | 6 | 27 | 46 |
| 21 | FormButton::void evaluateExtraParams() | 0.742 | 0.667 | 12 | 6 | 0.167 | 3 | 15 | 31 |
| 22 | ActionError::void evaluateExtraParams() | 0.846 | 0.750 | 6 | 2 | 0.000 | 0 | 8 | 14 |
| 23 | ActionMessage::void evaluateExtraParams() | 0.846 | 0.750 | 6 | 2 | 0.000 | 0 | 8 | 14 |
| 24 | Token::void evaluateExtraParams() | 0.679 | 0.750 | 9 | 3 | 0.000 | 0 | 12 | 21 |
| 25 | FieldError::void setFieldName() | 0.928 | 0.000 | 0 | 2 | 0.000 | 0 | 2 | 3 |
| 26 | Text::void addParameter() | 0.928 | 0.000 | 0 | 2 | 0.000 | 0 | 2 | 3 |
| 27 | Head::void evaluateParams() | 0.531 | 0.000 | 0 | 2 | 0.000 | 0 | 2 | 4 |

## 4.3 Extended Sorensen-Dice Index Results

The extended Sorensen-Dice index defined in Section 3.2 is greater than 0 when all three elements are included in a method structure. In the vector space model, the similarity is greater than 0 when at least one element of the three elements is included in a method structure. In other words, the extended Sorensen-Dice index requires the AND condition on the retrieval elements, while the vector space model requires the OR condition. Thus, the results of the extended Sorensen-Dice index are a subset of the results of the vector space model.

For example, the extended Sorensen-Dice index evaluated 0 for the *FieldError::void setFieldName()* method (No. 25 in Table 1) and the *Text::void addParameter()* method (No. 26 in Table 1), while the similarities obtained for these methods by the vector space model are 0.928. Both methods contain "*addParameter*" and "}"; however, these methods contain no *if*-statements.

Because "*addParameter*" is a rare term, the term weight for "*addParameter*" is so high that the similarity value works out to 0.928.

## 4.4. Derived Sequence Model Results

The DSRM's similarity is greater than 0 when the sequence [ *if{→addParameter→}* ] is included in an extracted method structure. This means that the DSRM imposes a more severe retrieval condition than the extended Sorensen-Dice model. In other words, the results of the DSRM are a subset of the results of the extended Sorensen-Dice model. The source code of *ActionError::void evaluateExtra Params()* (No. 22 in Table 1) is shown in Figure 4. The similarity of the method is estimated to be 0 by the derived sequence model because the method does not include the sequence [ *if{→addParameter→}* ]. Its similarity is 0.75 in the extended Sorensen-Dice model because the method includes two *if*-statements and two *addParameter* method calls.

A program is essentially represented by a sequence of statements. Because the DSRM computes the similarity based on a sequence of statements, it achieves higher performance than the other models.

```
protected void evaluateExtraParams() {
   boolean isEmptyList = true;
   Collection<String> actionMessages =  (List) findValue("actionErrors");
   if (actionMessages != null) {
      for (String message : actionMessages) {
         if (StringUtils.isNotBlank(message)) {
            isEmptyList = false;
            break;
         }
      }
   }
   addParameter("isEmptyList", isEmptyList);
   addParameter("escape", escape);
}
```

Figure 4. Example method that does not include any sequences of *if*-statements and *addParameter*

## 4.5. Summary of Experiments

Table 2 shows a summary of 27 retrieval experiments using the three models. Column three of Table 2 presents the number of methods retrieved by the DSRM with similarity values greater than 0. Column four presents the number of methods retrieved by the extended Sorensen-Dice model with similarity values greater than 0, and column five shows the number of methods retrieved by the vector space model with tf-idf weighting. The results of the experiment shown in Table 1 correspond to No. 14 in Table 2.

The degree of improvement of the DSRM over the extended Sorensen-Dice index is calculated by the following formula:

$$\frac{(No.methods\ by\ Sorensen-Dice)-(No.methods\ by\ DSRM)}{(No.methods\ by\ Sorensen-Dice)}$$

The degree of improvement of DSRM to the vector space model with tf-idf weighing is calculated by the similar formula.

The degree of improvement ranges from 0% to 90.1% over the extended Sorensen-Dice model, and ranges from 22.2% to 90.9% over the vector space model with tf-idf weighting. As previously mentioned, when the similarity is greater than 0, the results of the DSRM are a subset of the results of the extended Sorensen-Dice index, and the results of the extended Sorensen-Dice index are a subset of the results of the vector space model. Note that this set inclusion relationship is not always true when the top N-methods are selected. For example, for No. 23 and No. 27 in Table 2, the degree of improvement over the extended Sorensen-Dice model is 80.0%, and that of the vector space model is 60.9%. In these cases, the similarity of the vector space model with tf-idf weighting is 0.413, which is well above 0.

Table 2. Summary of 27 retrieval experiments

| No | Retrieval Condition | Derived Sequence Retrieval Model | Extended Sorensen -Dice Model | Vector Space Model with tf-idf | Degrees of Improvement | |
|---|---|---|---|---|---|---|
| | | | | | Extended Sorensen -Dice Model | Vector Space Model with tf-idf |
| 1 | if{ → addParameter | 21 | 24 | 27 | 12.5% | 22.2% |
| 2 | if{ → Logger.debug | 31 | 34 | 61 | 8.8% | 49.2% |
| 3 | if{ → Logger.warn | 32 | 35 | 43 | 8.6% | 25.6% |
| 4 | if{ → findValue | 6 | 16 | 33 | 62.5% | 81.8% |
| 5 | if{ → StringBuilder.append | 6 | 9 | 12 | 33.3% | 50.0% |
| 6 | if{ → findString | 4 | 11 | 20 | 63.6% | 80.0% |
| 7 | while{ → StringTokenizer.nextToken | 6 | 6 | 18 | 0.0% | 66.7% |
| 8 | while{ → Enumeration.nextElement | 5 | 5 | 11 | 0.0% | 54.5% |
| 9 | while{ → Iterator.next | 4 | 4 | 21 | 0.0% | 81.0% |
| 10 | for{ → iterator.next | 9 | 9 | 13 | 0.0% | 30.8% |
| 11 | for{ → List<String>.add | 4 | 5 | 6 | 20.0% | 33.3% |
| 12 | for{ → StringBuilder.append | 3 | 4 | 12 | 25.0% | 75.0% |
| 13 | catch{ → Logger.error | 18 | 25 | 43 | 28.0% | 58.1% |
| 14 | if{ → addParameter → } | 21 | 24 | 27 | 12.5% | 22.2% |
| 15 | if{ → Logger.debug → } | 26 | 34 | 68 | 23.5% | 61.8% |
| 16 | if{ → Logger.warn → } | 32 | 35 | 48 | 8.6% | 33.3% |
| 17 | if{ → findValue → } | 5 | 16 | 36 | 68.8% | 86.1% |
| 18 | if{ → StringBuilder.append → } | 4 | 9 | 11 | 55.6% | 63.6% |
| 19 | if{ → findString → } | 4 | 11 | 21 | 63.6% | 81.0% |
| 20 | if{ → } → } | 41 | 181 | 236 | 77.3% | 82.6% |
| 21 | if{ → } → if{ | 23 | 181 | 175 | 87.3% | 86.9% |
| 22 | if{ → } → else{ | 22 | 117 | 151 | 81.2% | 85.4% |
| 23 | for{ → if{ → } | 9 | 45 | 23 | 80.0% | 60.9% |
| 24 | if{ → addParameter → } → else{ | 4 | 7 | 19 | 42.9% | 78.9% |
| 25 | if{ → Logger.warn → } → } | 24 | 35 | 40 | 31.4% | 40.0% |
| 26 | if{ → } → } → } | 18 | 181 | 197 | 90.1% | 90.9% |
| 27 | for{ → if{ → } → } | 9 | 45 | 23 | 80.0% | 60.9% |
| | | | | **Average** | **39.4%** | **60.8%** |

Figure 5 shows a graph of the degree of improvement sorted by degree of improvement over the extended Sorensen-Dice model. The horizontal axis shows the sample number given in the first column of Table 2, and the vertical axis shows the degree of improvement in percentage. For all retrieval samples, the DSRM outperformed the extended Sorensen-Dice model except for samples No. 7, 8, 9, and 10. The extended Sorensen-Dice model is more successful than the vector space model with tf-idf weighting except for samples No. 23 and No. 27.

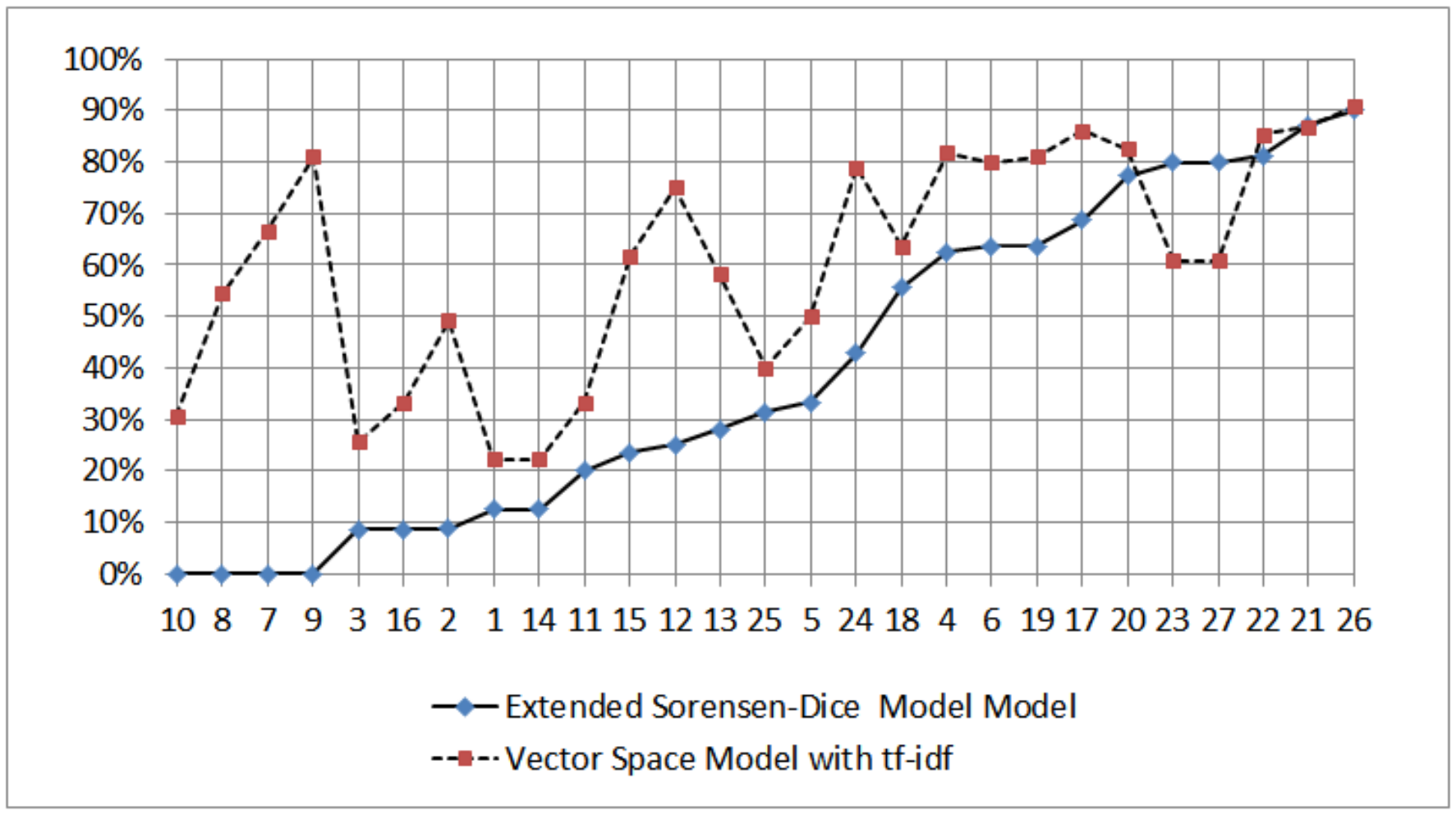

Figure 5. Degree of DSRM's improvement

## 5. Elapsed Time Comparisons

Table 3 summarizes the elapsed time in milliseconds of the three retrieval models for 27 sample retrievals. We measured the elapsed time using the following experimental environment:

CPU: Intel Core i3 540 3.07 GHz

Main memory: 4.00 GB

OS: Windows 7 64 Bit

The three retrieval models were implemented using Visual Basic for Excel 2010. The unique 1,420 statement fragments, including control statements and method calls, were extracted from the Struts 2.3.1.1 Core source code. Thus, a 1,420 × 2,667 matrix was stored in an Excel sheet for the retrieval experiments by the vector space model. All 2,667 methods were transformed into 2,667 sequences of extracted statements. They were also stored in an Excel sheet for the extended Sorensen-Dice model and the DSRM experiments. Through the experiments, all the data concerning retrieval were accessed from the Excel sheet cells. Thus, it is fair to say that the three retrieval model experiments were performed under equal conditions.

Table 3. Elapsed time of the three retrieval models (ms)

| No | Retrieval Sequence | Vector Space Model with tf-idf | Extended Sorensen -Dice Model Model | Derived Sequence Model |
|---|---|---|---|---|
| 1 | if{ → addParameter | 35618 | 2715 | 546 |
| 2 | if{ → Logger.debug | 35907 | 2933 | 499 |
| 3 | if{ → Logger.warn | 36458 | 3463 | 500 |
| 4 | if{ → findValue | 36037 | 2606 | 452 |
| 5 | if{ → StringBuilder.append | 33994 | 2917 | 500 |
| 6 | if{ → findString | 36442 | 2667 | 468 |
| 7 | while{ → StringTokenizer.nextToken | 35523 | 3260 | 515 |
| 8 | while{ → Enumeration.nextElement | 36444 | 3229 | 515 |
| 9 | while{ → Iterator.next | 36351 | 3214 | 500 |
| 10 | for{ → iterator.next | 35819 | 3167 | 515 |
| 11 | for{ → List<String>.add | 36388 | 3104 | 515 |
| 12 | for{ → StringBuilder.append | 37745 | 3042 | 514 |
| 13 | catch{ → Logger.error | 35322 | 3260 | 499 |
| 14 | if{ → addParameter → } | 36615 | 9162 | 1124 |
| 15 | if{ → Logger.debug → } | 35524 | 11138 | 1123 |
| 16 | if{ → Logger.warn → } | 36506 | 11809 | 1170 |
| 17 | if{ → findValue → } | 34776 | 9301 | 1155 |
| 18 | if{ → StringBuilder.append → } | 35713 | 10178 | 1108 |
| 19 | if{ → findString → } | 35682 | 9302 | 1108 |
| 20 | if{ → } → } | 36462 | 11545 | 1388 |
| 21 | if{ → } → if{ | 36164 | 10858 | 1388 |
| 22 | if{ → } → else{ | 36302 | 10873 | 1404 |
| 23 | for{ → if{ → } | 36520 | 11451 | 1466 |
| 24 | if{ → addParameter → } → else{ | 35069 | 16895 | 2824 |
| 25 | if{ → Logger.warn → } → } | 36411 | 18159 | 2839 |
| 26 | if{ → } → } → } | 35864 | 17877 | 2908 |
| 27 | for{ → if{ → } → } | 35257 | 18580 | 3034 |

Figure 6 shows a graph of the elapsed times presented in Table 3. The horizontal axis shows the sample number given in column one of Table 2, and the vertical axis shows the elapsed time in milliseconds. All 27 samples were processed in near-constant time in the vector space model because a given query is evaluated on the 1,420 × 2,667 matrix.

On the other hand, the extended Sorensen-Dice model and the DSRM require an elapsed time approximately proportional to the number of derived sequences related to a given retrieval condition. Both retrieval models generate two derived sequences for samples No. 1 to No. 13. As a result, three retrievals were executed. The average execution time was 0.171. For samples No. 14 to No. 23, both retrieval models executed seven retrievals. The average execution time was 0.187 milliseconds. For samples No. 24 to No. 27, both retrieval models executed 15 retrievals. The average execution time was 0.193 milliseconds. The elapsed time required for each derived sequence increases approximately 3%–8% due to the overhead involved in the retrieval process. The results in Table 3 indicate that the DSRM is approximately 10 times faster than the vector space model.

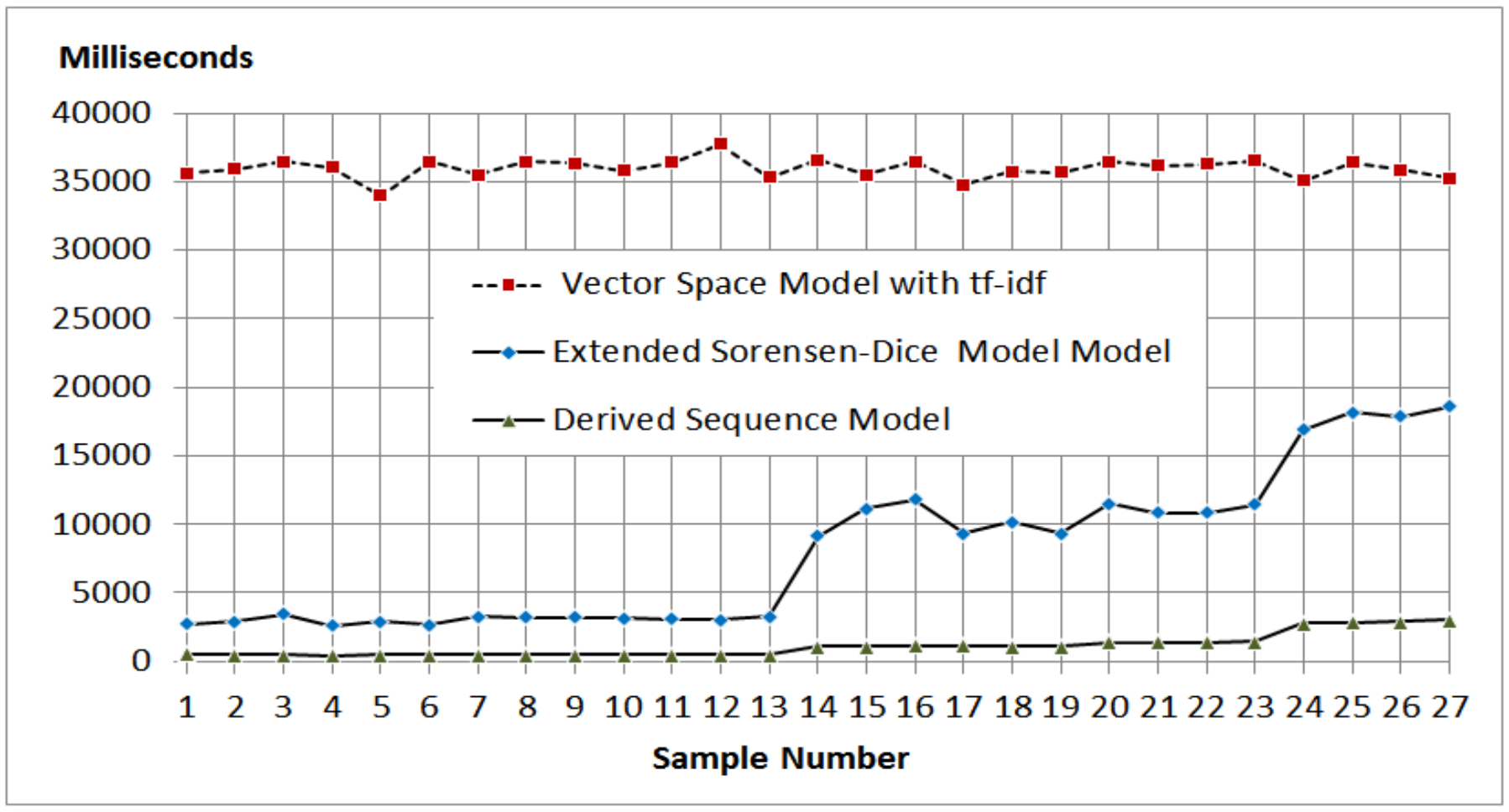

Figure 6. Elapsed time comparison

## 6. CONCLUSIONS

We presented a source code retrieval model that takes a sequence of statements as a retrieval condition. We conducted three types of experiments using the vector space model, the extended Sorensen-Dice model, and the derived sequence retrieval model (DSRM).

The key contribution of our approach is the definition of the DSRM's similarity measure as an extension of the Sorensen-Dice index and the evaluation of the DSRM's similarity measure on the Struts 2 Core source code, which is a moderate-sized Java program. The experimental results demonstrate that the DSRM's similarity measure shows higher selectivity than the other models, which is a natural consequence because a program is essentially a sequence of statements.

The results are promising enough to warrant further research. In future, we intend to improve our algorithms by combining information, such as the inheritance of a class and method overloading. We also plan to develop a better user interface, which would allow us to conduct further user studies and to more easily and precisely assess the retrieved code. In addition, we plan to conduct experiments using various types of open source programs available on the Internet.

## ACKNOWLEDGMENTS

We would like to thank Nobuhiro Kataoka, Tamotsu Noji, and Hisayuki Masui for their suggestions on engineering tasks to improve software quality.